\documentclass[a4paper,12pt]{article}
\pdfoutput=1
\usepackage{verbatim}
\usepackage{subfigure}
\usepackage{amsmath,graphicx,amssymb}
\usepackage{amsthm,bm}
\usepackage{multirow,color}
\usepackage[a4paper]{geometry}
\usepackage[affil-it]{authblk}
\begin{document}

\title{The infrared properties of the energy spectrum in freely decaying isotropic turbulence}
\markboth{\jobname}{\jobname}
\author{W.D. McComb}
\author{M.F. Linkmann}
\affil{SUPA, School of Physics and Astronomy, University of Edinburgh, UK}

\maketitle

\abstract 
The low wavenumber expansion of the energy spectrum takes the
well known form: \newline
$E(k,t)=E_2(t)k^2 + E_4(t)k^4 + \hdots$, where the
coefficients are weighted integrals against the correlation function
$C(r,t)$. We show that expressing $E(k,t)$ in terms of the longitudinal
correlation function $f(r,t)$ immediately yields $E_2(t)=0$ by cancellation.
We verify that the same result is obtained using the correlation function
$C(r,t)$, provided only that $f(r,t)$ falls off faster than $r^{-3}$ at
large values of $r$. As power-law forms are widely studied for the
purpose  of establishing bounds, we consider the family of model
correlations $f(r,t)=\alpha_n(t)r^{-n}$, for positive integer $n$, at
large values of the separation $r$. We find that for the special case
$n=3$, the relationship connecting $f(r,t)$ and $C(r,t)$ becomes
indeterminate, and (exceptionally) $E_2 \neq 0$, but that this solution
is unphysical in that the viscous term in the K\'{a}rm\'{a}n-Howarth equation
vanishes. Lastly, we show that $E_4(t)$ is independent of time, without
needing to assume the exponential decrease of correlation functions at large
distances.  

\section{Introduction}
\label{sec:context}

In this paper we restrict our attention to the three-dimensional,
isotropic turbulent motion of an incompressible fluid. We consider the
initial value problem posed by free decay from an arbitrarily chosen
energy spectrum. One interest in this work lies in its possible
relevance to the free decay of grid turbulence, a problem which has been
studied since the 1930s: see for example the classic books by Batchelor
\cite{Batchelor53}, Monin and Yaglom  \cite{Monin75}, and Hinze
\cite{Hinze75}. Another, increasing, interest lies in a practical
aspect; in that the choice of initial spectrum used in direct numerical
simulations of homogeneous isotropic turbulence  depends on which type
of infrared behaviour is allowed.

The problem posed by grid turbulence has, for many decades, proved to
be a rather controversial topic, with important issues remaining
unresolved to this day. Much of the disagreement centres on the
behaviour of the energy spectrum at low wavenumbers. However, recently
there has been growing doubt concerning the universal properties of the
large scales in this problem. It is increasingly believed
\cite{Valente12a, Meldi12} that the infrared properties of the energy
spectrum in isotropic turbulence depend on the initial conditions. That
is, they are not a universal feature of freely decaying grid turbulence.
Indeed, in a recent paper \cite{Vassilicos11} Vassilicos discussed the
possibility of an infinite number of invariants depending on the 
large-scale properties of the correlation function. However, we emphasise the
fact that in this paper we consider only the mathematical properties of
isotropic turbulence. The possible application to real situations
requires further attention and is a problem in its own right.

\section{Basic equations}
\label{sec:basic}

We summarise some standard equations and relationships here: for further
details see the books \cite{Batchelor53}, \cite{Monin75},
\cite{McComb14a}. Denoting the fluid velocity field by $u_n({\bf{x}},t)$,
where $n=1, 2, \mbox{or } 3$, we begin by defining the velocity covariance (or
pair-correlation) tensor, which we denote by $C_{nm}({\bf{r}},t)$, such that 
\begin{equation}
     C_{nm}({\bf{r}},t)=\langle u_n({\bf{x}},t)u_m({\bf{x}}+{\bf{r}},t) \rangle \ .
\end{equation}
It is well known that, from the symmetries of homogeneity and isotropy, we can
express this tensor through two scalar  functions $A(r,t)$ and $B(r,t)$,
which depend only on $r^2$, thus 
\begin{equation}
C_{nm}({\bf{r}},t)= A(r,t)r_n r_m + B(r,t)\delta_{nm} \ , 
\label{tensorCR}
\end{equation}
where $\delta_{nm}$ denotes the Kronecker delta. 
The correlation function $C(r,t)$ is defined as
\begin{equation}
C(r,t)\equiv \frac{1}{2}C_{nn}({\bf{r}},t)= \frac{A(r,t)r^2 + 3B(r,t)}{2} \ , 
\end{equation}
where we sum over repeated indices.
Since the flow we are investigating is taken to be incompressible,  the
velocity field must satisfy the continuity condition, which translates
to the correlation tensor as
\begin{equation}
  \frac{\partial C_{nm}({\bf{r}},t)}{\partial r_n} = 0  \ \ \ \forall \ {\bf{r}} \in \mathbb{R}^3 \ , 
\label{eq:incomp}
\end{equation}
where, again, we take repeated indices to be summed. 
This equation can be used to derive a further relation  
between the correlation function $C(r,t)$ and the functions $A(r,t)$ and $B(r,t)$
\begin{equation}
 C(r,t)= \frac{1}{2r^2}\frac{\partial}{\partial r}\big [ r^3[A(r,t)r^2 +B(r,t)]\big ] \ .   
\label{eq:CAB}
\end{equation}
The longitudinal correlation function $f(r,t)$ is defined as 
\begin{equation}
f(r,t)=\frac{C_{11}({\bf{r}},t)}{U^2(t)} \ ,
\end{equation}
where $U(t)$ is the rms value of any component of the velocity, and
we have chosen a set of basis vectors such that ${\bf{r}}=(r,0,0)$.
Therefore we obtain from \eqref{tensorCR} 
\begin{equation}
f(r,t)=\frac{A(r,t)r^2 +B(r,t)}{U^2(t)} \ ,
\end{equation}
and we can write \eqref{eq:CAB} in terms of $f(r,t)$ as 
\begin{equation}
 C(r,t)= \frac{U^2(t)}{2r^2}\frac{\partial}{\partial r}[r^3f(r,t)] \ .  
\label{eq:CAF}
\end{equation}

We may also introduce the triple correlation tensor:
\begin{equation}
C_{nm,l}({\bf{r}},t)=\langle u_n({\bf{x}},t)u_m({\bf{x}},t)u_l({\bf{x}}+{\bf{r}},t) \rangle  \ .
\end{equation}
Isotropy restricts the general form of this tensor to
\begin{equation}
 C_{nm,l}({\bf{r}},t) = A_1(r,t)r_n r_m r_l + A_2(r,t)(\delta_{ml}r_n + \delta_{nl}r_m) + A_3(r,t)\delta_{nm}r_l \ , 
\end{equation}
where the coefficients  $A_i(r,t)$ are functions of $r^2$.
Using the incompressibility condition, this equation can be expressed in terms of the triple correlation function $K(r,t)$
\begin{align}
 C_{nm,l}({\bf{r}},t) &= U^3(t)\left(K(r,t)-r\frac{\partial}{\partial r}K(r,t)\right)\frac{r_n r_m r_l}{2r^3} \nonumber \\
                      & + \frac{U^3(t)}{4r}\left(2K(r,t)+r\frac{\partial}{\partial r}K(r,t)\right)(\delta_{ml}r_n 
                        + \delta_{nl}r_m)-\frac{U^3(t)K(r,t)}{2r} \delta_{nm}r_l \ . 
\label{eq:tensor_K}
\end{align}
If we choose a set of basis vectors, as above, such that ${\bf{r}}=(r,0,0)$, we observe that
$C_{11,1}({\bf{r}},t)$ is a function of $r=|{\bf{r}}|$, and the triple correlation function
$K(r,t)$ may be written as
\begin{equation}
K(r,t)=\frac{C_{11,1}(r,t)}{U^3(t)} \ .
\end{equation}

For isotropic turbulence (see e.g. \cite{Batchelor53}, \cite{Monin75}, \cite{Hinze75})
the energy spectrum can be expressed in terms of the correlation
function $C(r,t)$ as
\begin{equation}
E(k,t) = \frac{2}{\pi} \int_0^{\infty} dr \ C(r,t) kr \sin{kr}  \ .
	\label{expansion}
\end{equation}
Expanding the sine in powers of $kr$ we obtain 
\begin{equation}
  E(k,t) = \frac{2}{\pi} \int_0^{\infty} dr \ C(r,t) \left [(kr)^2 - \frac{(kr)^4}{3!} +  \cdots \right ]  \ ,
\label{expansionC}
\end{equation}
which leads to the approximation of the energy spectrum at low wavenumbers by a Taylor polynomial of the form
\begin{equation}
E(k,t)= E_2(t) k^2 + E_4(t)k^4 + \cdots \ ,
\label{expansionC2}
\end{equation}
with 
\begin{equation}
  E_2(t) = \frac{2}{\pi}\int_0^{\infty} dr \ r^2 C(r,t) \ ,
\label{eq:E2}
\end{equation}
and  
\begin{equation}
  E_4(t) = -\frac{1}{3\pi}\int_0^{\infty} dr \ r^4 C(r,t) \ .
\end{equation}

\section{Observation that $E_2(t)=0$}
\label{sec:result}

Something interesting now happens if we express the energy spectrum in
terms of the longitudinal correlation function $f(r,t)$ instead of $C(r,t)$.
The expression for $E(k,t)$ becomes \cite{Batchelor53}, \cite{Monin75}, \cite{Hinze75},
\begin{equation}
  E(k,t) = \frac{U^2(t)}{\pi} \int_0^{\infty} dr \ f(r,t) kr (\sin{kr}-kr \cos{kr}) \ , 
\label{expansionF}
\end{equation} 
where the surface term that arises from the integration by parts (with respect to $r$) 
in the derivation of this equation from \eqref{expansion} and
\eqref{eq:CAF} vanishes for $f(r,t)$ decreasing faster than $r^{-2}$ as $r \to \infty$, 
which is assumed here. Expanding sine and cosine in powers of $kr$ then yields
\begin{equation}
  E(k,t) = \frac{U^2(t)}{\pi} \int_0^{\infty} dr \ f(r,t) kr \left[ (kr-kr) -\left( \frac{(kr)^3}{3!}-\frac{(kr)^3}{2!} \right ) +  \cdots \right ] \ , 
\label{E_f(r)}
\end{equation}
and we see that the lowest order term vanishes by cancellation. Hence
the first two coefficients in the expansion of $E(k,t)$ as given by
\eqref{expansionC2} are 
\begin{equation}
  E_2(t) =0 \   
\label{E20}
\end{equation}
and  
\begin{equation}
  E_4(t) = \frac{U^2(t)}{3\pi}\int_0^{\infty} dr \ r^4 f(r,t) \ .
\end{equation}
Thus equation \eqref{expansionF} forces $E_2(t)=0$, a result that we
have been unable to find in the literature.  

\subsection{Verification of $E_2(t)=0$}
\label{sec:verification}

One cannot immediately see (\emph{i.e.}~by inspection)
that $E_2(t)$ vanishes following equation \eqref{expansion}.  Both
\eqref{expansion} and \eqref{expansionF} are standard results for
isotropic turbulence describing the same quantity, therefore it should be
possible to verify $E_2(t)=0$ directly from \eqref{expansion}.

The coefficient $E_2(t)$, as given by \eqref{eq:E2}, can be expressed in terms of the
longitudinal correlation function by means of \eqref{eq:CAF}  as
\begin{align}
 E_2(t) &= \frac{2}{\pi}\int_0^{\infty} dr \ r^2  \frac{1}{2r^2}\frac{\partial}{\partial r}\left [ U^2(t)r^3f(r,t)\right ] \nonumber \\ 
     &= \frac{1}{\pi} \left [U^2(t)r^3f(r,t)\right ]_0^{\infty} \nonumber \\
     &= \lim_{r \to \infty} \frac{1}{\pi}U^2(t) r^3f(r,t) - \lim_{r \to 0} \frac{1}{\pi}U^2(t) r^3f(r,t) = 0 \ , 
\label{E2vanishes}
\end{align}
provided that $f(r,t)$ falls of faster than $r^{-3}$ at large $r$ and
that $f(r,t)$ is finite at $r=0$. This verifies the result $E_2(t)=0$
obtained from \eqref{expansionF}-\eqref{E20}. The large-scale behaviour
of $f(r,t)$ is further discussed in the next sections.

It is important to note that isotropy and incompressibility of the flow
are necessary for this derivation to hold. In general, a finite nonzero
value of $E_2(t)$ is only possible for an incompressible flow if
isotropy is broken. Hence, in numerical simulations of  homogeneous
isotropic turbulence of incompressible flows, initial spectra with an
infrared behaviour proportional to $k^2$ should not be used as this is
incompatible with isotropy. If the flow is compressible, $E_2(t)$ can be
non-zero for an isotropic system, as we cannot relate $f(r,t)$ to
$C(r,t)$ through \eqref{eq:CAF} in this case.

Therefore we conclude that for homogeneous isotropic turbulence of an
incompressible flow an expansion of $E(k,t)$ starts at $E_4(t)$,
provided that such an expansion is mathematically allowed. That is if
$E(k,t)$ is sufficiently many times differentiable. Hence we draw
attention to the fact that {\emph{a priori}} we have to assume the
finiteness of $E_4(t)$, and higher-order coefficients, in order to
obtain an expansion of $E(k,t)$ at small $k$. 

For the particular case $n=4$, it is well known that $E_4(t)$ can be
related to the integral of the angular momentum squared per
unit volume, provided this integral converges. 
The integral of the angular momentum squared per unit volume, that is the 
\emph{density} of the square of the angular momentum, should be finite in any 
physical system. 
Note that isotropic turbulence is, although somewhat artifical, still a
physical system and as such its properties must be described by
{\emph{physically admissible}} functions. A physically admissible function
is here defined as a function that yields
finite results for physical quantities which have the form of densities.
Thus on physical grounds it is reasonable to assume that the coefficient 
$E_4(t)$ is finite.

\section{Some properties of asymptotic power-law forms for the longitudinal
correlation coefficient}
\label{sec:modelfns}
In this subject it is usual to study power-law forms of the correlations
in order to establish various mathematical bounds. Accordingly, we
investigate the asymptotic properties of a model family of longitudinal
correlation functions of the form:
\begin{equation}
f(r,t)=\alpha_n(t)r^{-n},
\end{equation}
for positive integer $n$. We should emphasise two points. First, there
is no suggestion that correlations are actually of this form. That would
be highly unlikely, as there is no \emph{a priori} reason for the 
spatial and temporal dependences of $f(r,t)$ to be separable. 
Secondly, the functions $\alpha_n(t)$ must have
dimensions of length to the power $n$, in order to preserve the
dimensionless nature of $f(r,t)$. We now consider various properties of
these model forms.

\subsection{Indeterminacy in the differential relation between $C(r,t)$ and $f(r,t)$}
\label{sec:indeterminacy}

As indicated, we follow the usual practice in this subject and study the
power law decrease of $f(r,t)$ and $C(r,t)$ at the large scales. We note
that this is as a worst case, bearing in mind that $f(r,t)$ and $C(r,t)$
could decrease exponentially. We assume that $f(r,t)=\alpha_n(t)r^{-n}$
as $r \to \infty$, where $\alpha_n(t)=l_f(t)^{-n}$ for some time-dependent
length scale $l_f(t)$ in order to ensure that $f(r,t)$ is dimensionless,
as required. Then \eqref{eq:CAF} becomes
         \begin{equation}
           C(r,t) = \frac{U^2(t)\alpha_n(t)}{2}(3-n)r^{-n} \ ,  
         \end{equation}
which is (by inspection) indeterminate for $n=3$.

At the small scales \eqref{eq:CAF} is \emph{not} indeterminate for
$n\not =3$. But at the large scales the asymptotic behaviour of $f(r,t)$
is indeterminate for $n > 3$ due to a masking effect, as we shall now show. 

At the large scales the indeterminacy allows $f(r,t)$ to take the form 
         \begin{equation}
         f(r,t)=\alpha_3(t)r^{-3} + f_3(r,t) \ , 
         \label{eq:indet_f}
         \end{equation}
where $f_3(r,t)$ is a function that approaches zero faster than $r^{-3}$
as $r \to \infty$.  This result follows from direct substitution of
\eqref{eq:indet_f} into \eqref{eq:CAF}, whereupon the term
$\alpha_3(t)r^{-3}$ vanishes identically. The term $\alpha_3(t)r^{-3}$
masks the large-scale properties of $f_3(r,t)$.  Hence $C(r,t)$ can have
the same large-scale behaviour as $f_3(r,t)$ while $f(r,t)$ falls off as
$r^{-3}$.
 
We believe that the indeterminacy arises because $2C(r,t)$ is the trace
of the correlation tensor while $U^2(t)f(r,t)$ is defined as one of its
diagonal entries with respect to a suitable choice of basis vectors, the
other two diagonal entries being $U^2(t)g(r,t)$, where $g(r,t)$ is the
transverse correlation function \cite{McComb14a}. Hence
\begin{equation}
          2C(r,t)=U^2(t)\big[f(r,t)+2g(r,t)\big] \ ,
         \end{equation}
and the indeterminacy arises because there could be terms in $f(r,t)$
and $g(r,t)$ that vanish by cancellation. 
 
As we did in Section 3.1, we can use
\eqref{eq:CAF} to evaluate $E_2(t)$ directly by substituting it into
\eqref{eq:E2}:
\begin{equation}
          E_2(t) = \frac{2}{\pi}\int_0^{\infty} dr \ 
          r^2  \frac{1}{2r^2}\frac{\partial}{\partial r}\left [ U^2(t)r^3f(r,t)\right ]  
          = \frac{1}{\pi} \left [U^2(t)r^3f(r,t)\right ]_0^{\infty} 
          = \lim_{r \to \infty} \frac{1}{\pi}U^2(t) r^3f(r,t) \ . 
         \label{eq:E2vanishes}
         \end{equation}
From this we now conclude that $E_2(t)$ can only be nonzero if
$f(r,t)=\alpha_3(t)r^{-3}$ while $C(r,t)$ must decrease as fast as
$r^{-4}$ at the large scales to ensure convergence of the integral in
\eqref{eq:E2}. We next consider whether or not this special case is
compatible with the Navier-Stokes equation.

\subsection{Incompatibility of $f(r,t)=\alpha_3(t) r^{-3}$ with the Navier-Stokes equation}

We now consider the special case $f(r,t) = \alpha_3(t)r^{-3}$ at the
very large scales (where any contribution from terms that decrease 
faster than $r^{-3}$ is negligible), and ask if it is
compatible with the Navier-Stokes equation for freely decaying isotropic turbulence.  
We use the K\'{a}rm\'{a}n-Howarth equation (KHE), since it is
the evolution equation for the longitudinal correlation function which
is derived from the Navier-Stokes equation. This takes the form (see
\cite{McComb14a} pp.70-72):  
\begin{equation}
           \label{eq:KHE}
           \frac{\partial}{\partial t}U^2(t)f(r,t) = 2 \nu \left( \frac{4}{r}U^2(t)\frac{\partial}{\partial r}f(r,t)
           + U^2(t)\frac{\partial^2}{\partial r^2}f(r,t)\right) 
           + \frac{U^3(t)}{r^4} \frac{\partial}{\partial r}\big[r^4K(r,t)\big] \ ,
          \end{equation}
where $K(r,t)$ denotes the triple correlation function as defined in
Section 2. If we assume the particular behaviour
$f(r,t) = \alpha_3(t)r^{-3}$ at the large scales, \eqref{eq:KHE}
becomes
\begin{align}
       \label{eq:KHE_largesc}
       \frac{\partial}{\partial t}U^2(t)\alpha_3(t)r^{-3} &= 2 \nu \Big( \frac{4}{r}U^2(t)\alpha_3(t)\frac{\partial}{\partial r}
         r^{-3} +U^2(t)\alpha_3(t)\frac{\partial^2}{\partial r^2} r^{-3} \Big)
       + \frac{U^3(t)}{r^4} \frac{\partial}{\partial r}\big[r^4K(r,t)\big] \nonumber \\
       &=2 \nu \Big( -12U^2(t)\alpha_3(t)r^{-5} +12U^2(t)\alpha_3(t)r^{-5} \Big)
          + \frac{U^3(t)}{r^4} \frac{\partial}{\partial r}\big[r^4K(r,t)\big] \nonumber \\
        &=\frac{U^3(3)}{r^4} \frac{\partial}{\partial r}\big[r^4K(r,t)\big] \ ,
     \end{align}
where we see that the viscous term vanishes by cancellation. Hence for
the special case $f(r,t)=\alpha_3(t) r^{-3}$ the KHE does not contain a
viscous term, thus it ceases to describe freely decaying isotropic
turbulence. 

To be more precise, consider the differential equation we obtain from 
\eqref{eq:KHE_largesc}
\begin{equation}
  \label{eq:diffeq_K}
    \frac{\partial}{\partial t}U^2(t)\alpha_3(t)r = U^3(t) \frac{\partial}{\partial r}\big[r^4K(r,t)\big] \ ,
    \end{equation}
which yields after integration with respect to $r$, and some rearrangement,
\begin{equation}
   \label{eq:K_large}
   U^3(t) K(r,t) = \frac{1}{2}\frac{\partial}{\partial t}\big (U^2(t)\alpha_3(t)\big)r^{-2} + U^3(t)\beta(t)r^{-4} \ ,
   \end{equation}
where $\beta(t)=l_K(t)^{-4}$ for a time-dependent length scale $l_K(t)$ in order
to ensure that $K(r,t)$ is dimensionless. Batchelor and Proudman \cite{Batchelor56} derived 
$K(r,t) \sim r^{-4}$ as a bound on the large-scale behaviour of $K(r,t)$, which implies  
that the term proportional to $r^{-2}$ in \eqref{eq:K_large} vanishes, rendering 
$U^2(t) \alpha_3(t)$ time independent. 

Thus the special case $f(r,t)=\alpha_3(t) r^{-3}$ implies that the longitudinal velocity 
correlations at the large scales are time-independent, since
\begin{equation} 
\langle u_L({\bf{x}},t)u_L({\bf{x}}+{\bf{r}},t)\rangle = U^2(t)f(r,t)=U^2(t)\alpha(t)r^{-3} \ ,  
\end{equation}
and $U^2(t)\alpha(t)=const$. This is not compatible with the Navier-Stokes equation 
describing freely decaying turbulence, as it implies the persistence of large-scale 
velocity correlations at all times, which is only possible if $u_L({\bf{x}},t)$ and
$u_L({\bf{x}}+{\bf{r}},t)$ do not decay over time. We thus conclude that the special 
case $f(r,t)=\alpha_3(t) r^{-3}$ is unphysical.

\section{$E_4(t)$ as an integral invariant}
\label{sec:dE4/dt=0}
As we have seen in Section 3, the coefficient $E_2(t)$ in the expansion of the energy spectrum vanishes, 
provided $f(r,t)$ is sufficiently well behaved, which it must be, as we have shown in the last section. 
We now consider the question of whether or not the next higher-order coefficient $E_4(t)$ depends on time. 
To investigate the time dependence of $E_4(t)$ we study the triple correlation tensor, 
as defined in Section 3. 

Assuming its existence (as is usual practice in this field), we denote
 by $T_{nm,l}({\bf{k}},t)$ the Fourier transform of the correlation tensor
 which we write explicitly (see Monin and Yaglom \cite{Monin75} p.~71 eq.~$(12.140)$):
\begin{equation}
 T_{nm,l}({\bf{k}},t) = iF_3(k,t)\left (\delta_{ml}\frac{k_n}{k} + \delta_{nl}\frac{k_m}{k} -\frac{2k_n k_m k_l}{k^3} \right ) \ , 
\label{eq:tensor_F}
\end{equation}
where $F_3(k,t)$ is a scalar function. 
As in Monin and Yaglom \cite{Monin75} p.~73, the transfer spectrum $T(k,t)$ is related to $F_3(k,t)$ by
\begin{equation}
T(k,t)=-8\pi k^3 F_3(k,t) \ . 
\end{equation}
The transfer spectrum can be expressed in terms of the triple correlation function as
\begin{equation}
 T(k,t)= -\frac{U^3}{\pi} \int_0^{\infty} dr \ \frac{K(r,t)}{r} \big[\{(kr)^2-3\} kr \sin{kr} +3(kr)^2\cos{kr} \big] \ ,
\label{eq:T}
\end{equation}
where we have made use of Batchelor and Proudman's bound on the large-scale behaviour of $K(r,t)$ \cite{Batchelor56}, 
which ensures that the surface term that arises from the integration by parts in the derivation of this 
equation vanishes.
(Note that   
there is a $-$ve sign missing in Monin and Yaglom's equation $(12.142''')$.)
Expanding the trigonometric functions in powers of $kr$ yields
\begin{align}
 T(k,t)&=- \frac{U^3(t)}{\pi} \int_0^{\infty} dr \ \frac{K(r,t)}{r} \left ((kr)^4-3(kr)^2+3(kr)^2 - \left [\frac{(kr)^6-3(kr)^4}{3!}+\frac{3(kr)^4}{2!} \right] + \hdots \right )  \nonumber \\ 
       &= \frac{U^3(t)}{\pi} \int_0^{\infty} dr \ \left[ \frac{r^5}{15}k^6 -\frac{r^7}{210}k^8 + \hdots \right ]K(r,t) \ .
\label{eq:expansionT}  
\end{align}
(We note that this expansion had previously been derived by Yoffe 
\cite{YoffePC12}.)
We see immediately that the terms proportional to $k^2$ and $k^4$ vanish by 
cancellation. 

The time invariance of $E_4(t)$ follows now from energy conservation. In
isotropic turbulence, energy conservation is expressed through the energy 
balance equation 
\begin{equation}
T(k,t)=\left( \frac{\partial}{\partial t} + 2\nu k^2\right)E(k,t) \ , 
\label{Lin}
\end{equation}
which is also referred to as the Lin equation in the literature \cite{McComb14a}.
Given the Taylor expansion of $E(k,t)$ as obtained in Section 3, the energy balance equation in the infrared can be written as
\begin{equation}
T(k,t)=\left( \frac{\partial}{\partial t} + 2\nu k^2 \right)E(k,t)=\frac{\partial}{\partial t}E_4(t)k^4 + \mathcal{O}(k^5)  \ , 
\label{Lin_expansion}
\end{equation}
which gives a small-wavenumber approximation of the transfer spectrum
\begin{equation}
T(k,t)=A_4k^4 + {\mathcal{O}}(k^5)  \ ,
\label{T_expansion}
\end{equation}
with coefficent $A_4 = \partial E_4(t)/ \partial t$. The viscous terms enter at higher order in $k$. 
Since
$E(k,t)$ and $T(k,t)$ are related through the energy balance \eqref{Lin}, we have found a polynomial of order four that approximates $T(k,t)$ for small $k$.
Now recall that the Taylor polynomial of a function is unique, hence the polynomial we found is the Taylor polynomial of $T(k,t)$. 
Therefore $T(k,t)$ is at least four times differentiable at zero with respect to $k$ and the coefficient $A_4$ reads 
\begin{equation}
A_4(t)=\frac{\partial^4}{\partial k^4} \frac{T(k,t)}{4!} \ .
\end{equation} 
We see directly from \eqref{eq:expansionT} that all derivatives of $T(k,t)$ with respect to $k$ which are of order $n\leqslant 5$ vanish, hence
\begin{equation} 
\frac{\partial}{\partial t}E_4=A_4 = 0 \ ,
\end{equation} 
and the time derivative of Loitsyansky's integral $E_4$ vanishes. 

\section{Conclusion}

Studying the spectral representations of the correlation tensors and
their corresponding correlation functions we have shown that the lowest-order 
term in an expansion of the energy spectrum at low wavenumbers is
proportional to $k^4$. 

It should be noted that the vanishing of the coefficient $E_2$ is purely
a property of the Fourier weight function and requires only that the
longitudinal correlation function $f(r,t)$ is reasonably well behaved.
For the result to hold when working with the full correlation function
$C(r,t)$, we found the more specific requirement that $f(r,t)$ should
fall off faster than $r^{-3}$ at large values of $r$.

The indeterminacy in the relationship between $C(r,t)$ and $f(r,t)$, as
given by \eqref{eq:CAF} for the case of $r^{-3}$, was found to be
unphysical because this particular power law is incompatible with the
Navier-Stokes equations.

Furthermore we have recovered what is known in the literature as
Loitsyansky's invariant.  That is we have shown that the $k^4$ infrared
behaviour of the energy spectrum persists over time, without the usual
necessity to assume exponential decrease of either the correlation or
the triple-correlation functions with distance.  In short, homogeneous
isotropic turbulence is of Loitsyansky type and this is a universal
feature of it.

\section*{Acknowledgements}

The authors thank Samuel Yoffe and Arjun Berera for reading through 
previous drafts of this work and for many helpful discussions and suggestions.  
M. L. acknowledges financial support by the UK Engineering and Physical Sciences
Research Council (EPSRC).

\bibliographystyle{unsrt}
\bibliography{wdm}	 

\end{document}